\documentclass[notitlepage,showpacs,aps,amsmath,amssymb,unsortedaddress,superscriptaddress]{revtex4-1}
\usepackage{bm,graphicx,color}
\begin{document}
\title{Three-dimensional light bullets in a Bragg medium with carbon
  nanotubes}
\author{Alexander V. Zhukov} \affiliation{Singapore University of Technology
  \& Design, 8 Somapah Road, 487372 Singapore}\email{alex.zhukov@outlook.sg}
\author{Roland Bouffanais} \affiliation{Singapore University of Technology \&
  Design, 8 Somapah Road, 487372 Singapore}
\author{Mikhail B. Belonenko} \affiliation{Laboratory of Nanotechnology,
  Volgograd Institute of Business, 400048 Volgograd, Russia}
\affiliation{Volgograd State University, 400062 Volgograd, Russia}
\author{Ilya S. Dvuzhilov} \affiliation{Volgograd State University, 400062
  Volgograd, Russia}
\author{Yulia V. Nevzorova} \affiliation{Volgograd State University, 400062
  Volgograd, Russia} \date{\today}
\begin{abstract}
  We present a theoretical study of the propagation of three-dimensional
  extremely short electromagnetic pulses (a.k.a. light bullets) through a
  Bragg medium containing an immersed array of carbon nanotubes. We
  demonstrate the possible stable propagation of such light bullets. In
  particular, our results suggest these light bullets can carry information
  about the Bragg medium itself.
\end{abstract}
\pacs{78.67.-n 78.66.Tr 78.67.Ch 78.70.Gq}
\maketitle
\section{Introduction}
% --------------------------

%
Among the vast breadth of nonlinear optical phenomena, the propagation of
extremely short three-dimensional (3D) spatiotemporal optical solitons has
attracted considerable attention given the range of different properties
observed with various nonlinear media~\cite{0,1,M1,M2,M3,2}. Localized
electromagnetic wave packet inevitably spreads both in space and time under
the concurrent effects of dispersion and diffraction present in any
medium. Significant research activity has been dedicated to devising new ways
to overcome these universal broadening effects in order to generate sustained
localized wave packets~\cite{0,1,M1,M2,M3,2}.  Such traveling wave packets
that are localized while retaining their spatiotemporal shape---in spite of
diffraction and dispersion effects---are referred to as ``light
bullets''. When propagating through a nonlinear medium, three-dimensional (3D)
light bullets tend to vanish as a consequence of a host of
instabilities~\cite{M1}. One can say that a light bullet is a natural
generalization of well-known one-dimensional (1D) electromagnetic
solitons~\cite{3} to the greater dimension case. In particular, the
distinguishing feature of these extremely short pulses---interesting and
relevant from several viewpoints, including the technological one---is that it
is impossible to make a partition of the form of the electromagnetic pulse
between its envelope and its carrier part. As a result, the well-established
method of multiscale expansion is no longer suitable for solving Maxwell's
equations. The latter must be solved without discarding any
derivatives~\cite{4,5}. Note that in general, the medium produces unavoidable
dispersive effects. Therefore, the appearance of solutions with localized
energy requires consideration of nonlinear effects even in the one-dimensional
case.

Nanoscale carbon materials such as carbon nanotubes (CNTs) have strongly
pronounced nonlinear properties in the optical range, which make them very
attractive for a vast range of uses in applications~\cite{6,7,8}.  They can
also be used as the medium in which the formation of light bullets
occurs. Corresponding calculations were made for one-, two- and
three-dimensional cases~\cite{9,10,11,12,13,Z1,Z2,Z3} and it was found that
the CNT medium allows for the possible stable propagation of light bullets.
There is only one rather weighty constraint that comes from practical
considerations: the speed of propagation of light bullets is only defined by
the refractive index of the medium and it can only be tuned in a fairly narrow
interval of values. One possible solution to this issue is to consider the
additional modulation of the refractive index of the medium, forming the
so-called Bragg environment. These considerations were first successfully
applied to a two-dimensional case in Ref.~\cite{Z4}, while here we address the
more challenging and realistic problem associated with the 3D case.

In the case of a Bragg environment, the propagation speed of the wave packet
is defined by both the period and depth of modulation of the refractive index,
due to partial reflection and the further interference of the wave packet. In
this case, it is theoretically possible to control the speed of light bullets
in such an environment. In practice, the refractive index modulation is
possible with the use of an external DC field in any environment that allows
for either Kerr or Faraday effects, and which contains CNTs.  It is worth
adding that the light bullets have recently been reported in pure Kerr
media~\cite{ad1, ad2}.  Note that the simple considerations given in
Ref.~\cite{10} to the existence of light bullet do not apply in the present
case owing to the fact that there is no translational invariance. It is also
quite obvious that the additional variance introduced by the Bragg environment
will not lead to the collapse of the light bullets. The importance of
practical applications and the considerations set out above provide the
impetus for the present study.

\section{Fundamental equations}
% --------------------------

%
In our study, we use the strong-coupling approximation for the electronic
structure of CNTs in the framework of the analysis of the dynamics of
$\pi$-electrons. The dispersion relation for a zigzag-type CNT $(m, 0)$ reads
as~\cite{7}
\begin{equation}
  E_s({\bf p}) = \pm\gamma \left\{ 1+ 4\cos (ap_z)\cos(\pi s/m)+ 4\cos^2 (\pi s/m)\right\}^{1/2},
  \label{1}
\end{equation}
where $\gamma = 2.7$ eV, $a = 3b/2\hbar$, and $b = 0.152$ nm is the distance
between neighboring carbon atoms. Note that the quasimomentum ${\bf p}$ is
represented here as $(p_z, s), s = 1,2,...m$.

\begin{figure}
  \includegraphics[width=16cm]{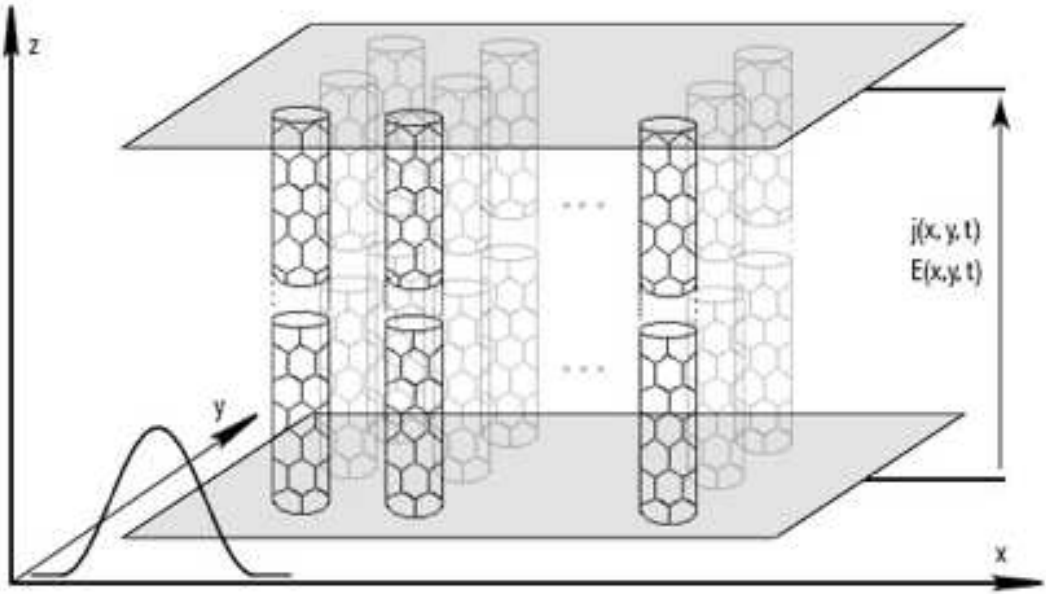}
  \caption{\label{fig1} Geometry of the problem with carbon nanotubes aligned
    along the $z$-direction and contributing to generating a Bragg grating
    along the $x$-direction.}
\end{figure}

The system consists of an ordered array of CNTs embedded into any medium whose
refractive index is harmonic along the bullet propagation axis. The variations
of the refractive index of this medium are fully characterized by their period
and magnitude. When constructing a model of the ultrashort optical pulse
propagation in a Bragg environment with nanotube system, where the geometry is
shown in Fig.~\ref{fig1}, we will describe the electromagnetic pulse field on
the basis of Maxwell's equations in the Coulomb gauge~\cite{14}, namely ${\bf
  E} = - {\partial {\bf A}}/{c \partial t}$.
The vector-potential is thereby expected to take the reduced form ${\bf A} =
\left( 0, 0, A_z(x,y,z,t)\right)$. The governing propagation equation can be
written as
\begin{equation}
  \frac{\partial^2 {\bf A}}{\partial x^2} + \frac{\partial^2 {\bf A}}{\partial y^2} 
  + \frac{\partial^2 {\bf A}}{\partial z^2} - \frac{n^2(x)}{c^2}\frac{\partial^2 {\bf A}}{\partial t^2}
  + \frac{4\pi}{c} {\bf j} = 0,
  \label{2}
\end{equation}
where $n(x)$ represents the spatial variations of the refractive index along
the $x$-axis, i.e., the 1D Bragg grating; ${\bf j}$ is the current density
that originates from the influence of the electric field pulse onto the
electrons in the conduction band of the CNTs.  Here, we neglect the
diffraction spreading of the laser beam in the direction along the axis of the
nanotubes. The electric field eventually induced by the substrate itself is
not considered in this proposed formalism. We also discard possible inter-gap
jumps, which results in restricting the possible frequencies of laser pulses
to the near-infrared region.  The typical size of CNTs and the distance
between them are both much smaller than the spatial scale of the region where
ultrashort pulses are localized. This means that we can appropriately work
under the continuous-medium approximation, and thus consider the current
density as homogeneously distributed over a given volume. The characteristic
length scale associated with spatial variations of the refractive index in the
Bragg medium is even larger, and therefore does introduce any further
restriction to our modeling framework.

Analytical evaluation of the current density ${\bf j}$ is essentially similar
to that provided in Ref.~\cite{Z4}. We however provide it here briefly for the
sake of self-consistency.  Typical relaxation time for electrons in CNTs can
be estimated at $3\times 10^{-13}$ s, then the electron ensemble (on time
scales of the order of $10^{-14}$ s, which is typical for ultrashort EM
pulses) can be described by the collision-less Boltzmann equation,
\begin{equation}
  \frac{\partial f_s}{\partial t}  - \frac{q}{c}\frac{\partial A_z}{\partial t} \frac{\partial f_s}{\partial p_z}  = 0,
  \label{3}
\end{equation}
where $f_s = f (p_z,s,t)$ is the electron distribution function, which
implicitly depends on the spatial coordinates; $q$ is the electron charge, and
$c$ is the speed of light in vacuum. At the initial instant, the distribution
$f$ is classically given by the equilibrium Fermi--Dirac distribution
\begin{equation}
  f_{s0} = \left\{ 1 + \exp\left( E_s({\bf p})/k_BT\right)\right\}^{-1},
  \nonumber
\end{equation}
where $T$ is the temperature, and $k_B$ is the Boltzmann constant. The current
density ${\bf j} = (0,0,j_z)$ is given by
\begin{equation}
  j_z = \frac{q}{\pi\hbar} \sum_s \int f_s(p_z) v_z \,dp_z ,
  \label{4}
\end{equation}
where we have introduced the group velocity $v_{sz} = \partial E_s({\bf
  p})/\partial p_z$. Solving Eq.~\eqref{3} by means of the method of
characteristics allows us to obtain
\begin{equation}
  j_z = \frac{q}{\pi\hbar} \sum_s \int_{-q_0}^{q_0} dp_z v_{sz} \left( p - \frac{q}{c}A_z\right) f_0({\bf p}).
  \label{5}
\end{equation}
The integration in Eq.~\eqref{5} is performed over the first Brillouin zone
with $q_0 = 2\pi\hbar/3b$. The group velocity can conveniently be expanded as
a Fourier series,
\begin{equation}
  v_{sz}(x) = \sum_m a_{ms} \sin (mx),
  \nonumber
\end{equation}
where
\begin{equation}
  a_{ms} = \frac{1}{\pi} \int_{-\pi}^{\pi} v_{sz} (x) \sin(mx)\, dx.
  \nonumber
\end{equation}
The propagation equation for the vector potential becomes
\begin{equation}
  \frac{\partial^2 A_z}{\partial x^2} + \frac{\partial^2 A_z}{\partial y^2} + \frac{\partial^2 A_z}{\partial z^2}
  - \frac{n^2(x)}{c^2}\frac{\partial^2 A_z}{\partial t^2}
  + \frac{q}{\pi\hbar} \sum_{m}c_m\sin\left( \frac{maq}{c} A_z\right) = 0,
  \label{6}
\end{equation}
where
\begin{equation}
  c_m = \sum_m a_{ms} b_{ms}, \qquad \text{with} \quad b_{ms} = \int_{-q_0}^{q_0} dp_z \cos(map_z) f_0 ({\bf p}).
  \nonumber
\end{equation}
At this stage, for computational convenience, we pass to the cylindrical
coordinates, where Eq.~\eqref{6} reads
\begin{equation}
  \frac{\partial^2 A_z}{\partial x^2} + \frac{1}{r}\frac{\partial}{\partial r}\left(r\frac{\partial A_z}{\partial r}\right) 
  - \frac{n^2(x)}{c^2}\frac{\partial^2 A_z}{\partial t^2}
  + \frac{q}{\pi\hbar} \sum_{m}c_m\sin\left( \frac{maq}{c} A_z\right) = 0,
  \label{7}
\end{equation}
where $r = \sqrt{y^2+z^2}$.

Our estimations show that the coefficients $c_m$ decrease with increasing $m$
approximately as $(1/2)^m$. Then---when computing Eq.~\eqref{6}---we can
restrict ourselves to the first ten terms, and subsequently increase the
number of terms depending on the required accuracy.

\section{Results and discussion}
% ----------------------------------------------------------------------------------------------

%
For the numerical solution of Eq.~\eqref{7}, we have implemented an explicit
finite-difference scheme for hyperbolic equations~\cite{17}. The step sizes,
both in time and space, were iteratively decreased by a factor of two, until
the obtained solution became unchanged to the eighth decimal place. Initial
conditions for the vector potential have been chosen to have the following
form:
\begin{align}
  A_z(x,r,t=0)&= A_0 \exp \left\{ - \frac{x^2}{\gamma^2}\right\}
  \exp\left\{-\frac{r^2}{\beta^2}\right\},
  \label{8}\\
  \frac{dA_z}{dt}\bigg\vert_{t=0} &= \frac{2vx}{\gamma^2}A_0 \exp \left\{
    -\frac{x^2}{\gamma^2} \right\} \exp\left\{-\frac{r^2}{\beta^2}\right\},
  \label{9}
\end{align}
where $v$ is the initial pulse velocity, $\beta$ and $\gamma$ are the
parameters determining the pulse width along $r$ and $x$, respectively. The
refractive index of the medium has been modeled as $n(x) = n_0 \left( 1+
  \alpha (2\pi x/\chi )\right)$, where $\alpha$ is the modulation depth, and
$\chi$ is the period of the Bragg grating, where we have taken $\alpha =
0.05$.

\begin{figure}
  \includegraphics[width=16cm]{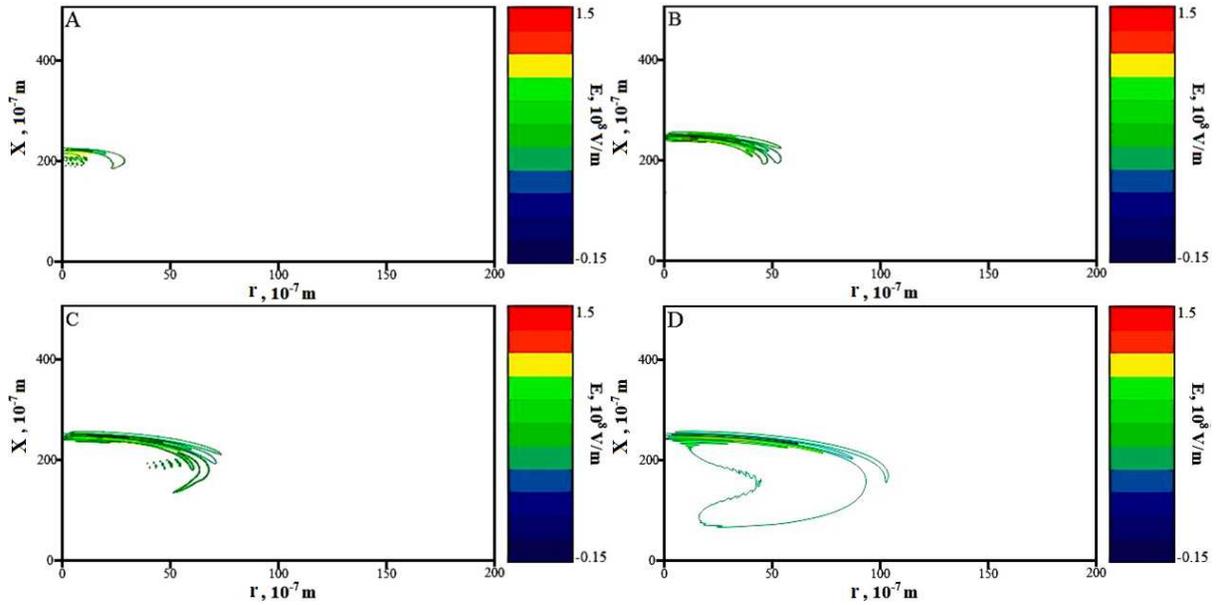}
  \caption{\label{fig2} Propagation of a light bullet in the Bragg environment
    (lattice period $\chi = 2.5$ $\mu$m) with carbon nanotubes at given
    instants of time ($T_0=2.5$ ps): (A) $T = T_0$; (B) $T = 2T_0$; (C) $T =
    3T_0$; (D) $T = 4T_0$.  Values of the field intensity are mapped on a
    color scale, where the maximum values correspond to red and the minimum
    ones to dark blue.}
\end{figure}

The results of numerical calculations show that the propagation of the light
bullet is stable and the resulting evolution is shown in Fig.~\ref{fig2}.

As can be seen from the resulting dependencies, the light bullet in a Bragg
environment with cross-modulation of the refractive index is not experiencing
any considerable broadening, but there are energy fluctuations in the
environment after its passage. We attribute this to the lack of balance
between the environmental dispersion and nonlinearity of the medium (in
contrast to the case of solitons), resulting thereby in a light bullet shape
changes. We also note that, despite the change in the shape of the light
bullet, its energy is still concentrated in a limited area.

\begin{figure}
  \includegraphics[width=16cm]{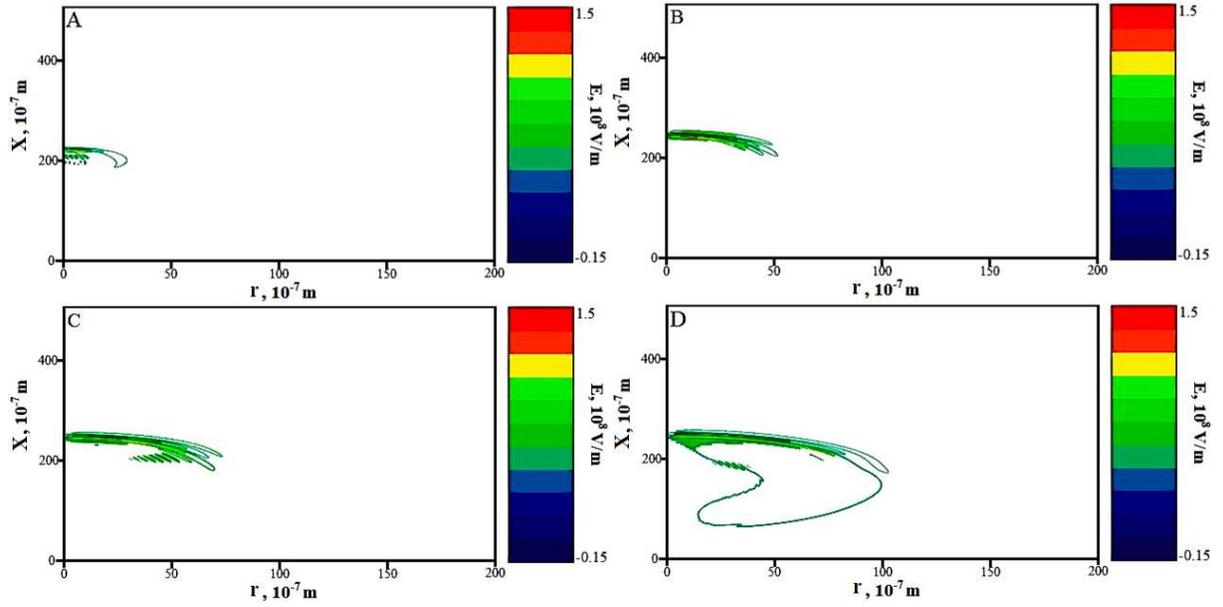}
  \caption{\label{fig3} Propagation of a light bullet in the Bragg environment
    (lattice period $\chi = 5$ $\mu$m) with carbon nanotubes at given instants
    of time ($T_0=2.5$ ps): (A) $T = T_0$; (B) $T = 2T_0$; (C) $T = 3T_0$; (D)
    $T = 4T_0$.  Values of the field intensity are mapped on a color scale,
    where the maximum values correspond to red and the minimum ones to dark
    blue.}
\end{figure}

\begin{figure}
  \includegraphics[width=16cm]{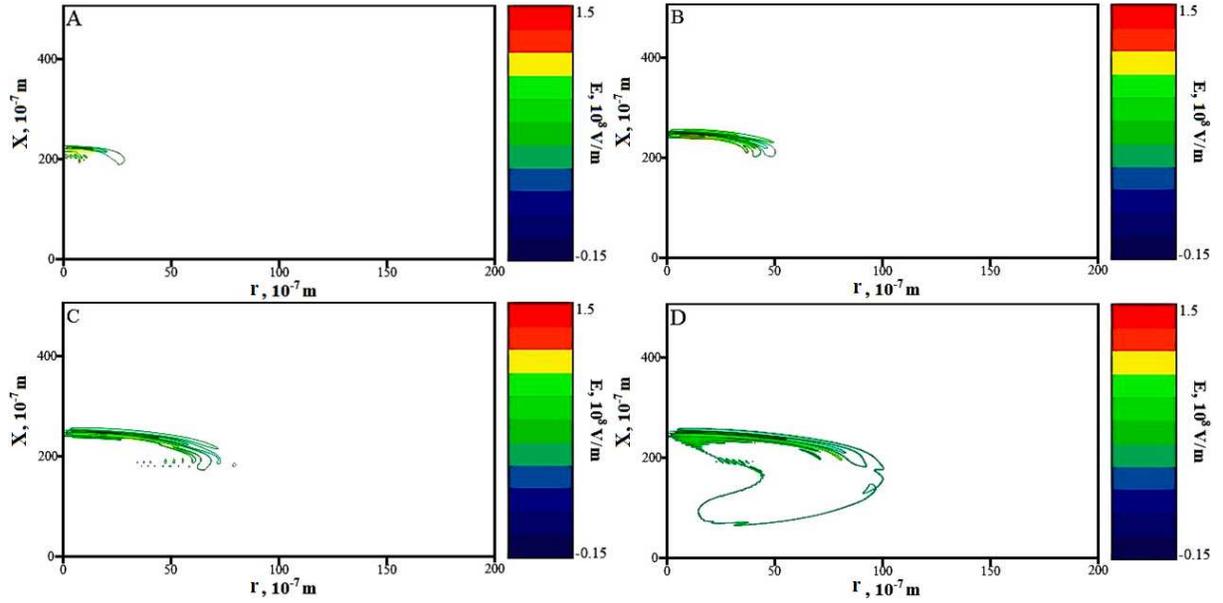}
  \caption{\label{fig4} Propagation of a light bullet in the Bragg environment
    (lattice period $\chi = 10$ $\mu$m) with carbon nanotubes at given
    instants of time ($T_0=2.5$ ps): (A) $T = T_0$; (B) $T = 2T_0$; (C) $T =
    3T_0$; (D) $T = 4T_0$.  Values of the field intensity are mapped on a
    color scale, where the maximum values correspond to red and the minimum
    ones to dark blue.}
\end{figure}

To support our qualitative conclusions, we have performed the same numerical
analysis for two and four times larger Bragg periods $\chi$, illustrated in
Figures \ref{fig3} and \ref{fig4}, respectively.

As can be seen from the evolution of the propagation of three-dimensional
extremely short optical pulses (Figures \ref{fig2}--\ref{fig4}), the light
bullet somewhat changes its configuration and spreading of shape inevitably
occurs over time due to the unavoidable dispersive effects of the medium. By
extremely short pulses, we refer to light bullets with spatial extent of the
order of $10^{-5}$~m and temporal duration of the order of
$10^{-13}$~s. These pulses correspond to a few electric field oscillations
(less than 10). The solution for three-dimensional light bullets in Bragg
environment with carbon nanotubes remains localized, but changes its spatial
structure due to the lateral dispersion. The combined effect of the pulse
spreading due to the dispersion and nonlinearity leads to the formation of a
multi-peak transverse structure, which nevertheless remains localized in a
bounded spatial domain.

Thus, we demonstrated the possibility for the stable propagation of
three-dimensional light bullets through an array of CNTs immersed in a medium
with periodically varying refractive index.  Note that from a practical point
of view, this result is important because it allows one to control the speed
of light bullets by varying the parameters of the Bragg medium---the Bragg
period $\chi$. At the same time, the propagation of light bullets in a Bragg
medium has some significant differences from the case of a medium with a
constant refractive index. Perhaps the most important difference is that the
light bullets in a Bragg environment have a more complex transverse structure,
which we believe is related to the excitation of internal modes of light
bullets resulting from the interaction with the inhomogeneity of the
refractive index of the medium. Previously, such an effect has already been
observed in the solution of the problem of the interaction of ultrashort
pulses with metallic heterogeneity in environments with CNTs~\cite{18}.

\section*{Acknowledgments}

A. V. Zhukov and R. Bouffanais are financially supported by the SUTD-MIT
International Design Centre (IDC).

% -------------------------------------------------------------------------------------------------------------

%

%
\end{document}